\documentclass[aps,twocolumn,groupedaddress,showpacs,amssymb,prb,floatfix,preprintnumbers]{revtex4}

\usepackage{graphicx,color}
\usepackage{subfigure}
\usepackage{verbatim}

\begin{document}
	\thispagestyle{myheadings}
	
	\title{Magnetic and electronic phases of U$_\mathbf{2}$Rh$_\mathbf{3}$Si$_\mathbf{5}$}
	
	\author{J. Willwater$^1$}
	\author{N. Steinki$^1$}
	\author{R. Reuter$^1$}
	\author{D. Menzel$^1$}
	\author{H. Amitsuka$^2$}
	\author{V. Sechovsky$^3$}
	\author{M. Valiska$^3$}
	\author{M. Jaime$^4$\footnote{Present Address: Physikalisch-Technische Bundesanstalt, Bundesallee 100, 38116 Braunschweig}}
	\author{F. Weickert$^5$\footnotemark[1]}
	\author{S. S\"ullow$^1$}
	\affiliation{
	$^1$Institut f\"ur Physik der Kondensierten Materie, TU Braunschweig, D-38106 Braunschweig, Germany
	\\
	$^2$Department of Physics, Hokkaido University, Sapporo 060-0810, Japan 
	\\
	$^3$Faculty of Mathematics and Physics, Charles University, Ke Karlovu 5, Prague 2, Czech Republic
	\\
	$^4$National High Magnetic Field Laboratory, Los Alamos National Laboratory, Los Alamos, New Mexico 87545, USA
	\\
	$^5$NHMFL, Florida State University, Tallahassee, FL 32310, United States	
	}
	
	\date{\today}
	
	\begin{abstract}
		We present a detailed study of the magnetic and electronic properties of $\mathrm{U}_2\mathrm{Rh}_3\mathrm{Si}_5$, a material that has been demonstrated to exhibit a first order antiferromagnetic phase transition. From a high magnetic field study, together with extensive experiments in moderate fields, we establish the magnetic phase diagrams for all crystallographic directions. The possibility of an electronic phase in a narrow interval above the N\'{e}el temperature as a precursor of a magnetic phase is discussed.
	\end{abstract}
	
	 \maketitle
	
	\section{\label{sec:level1}Introduction}
	For many years now, the variety of exotic ground states in intermetallic uranium compounds has been the focus of extensive research efforts. Special attention was given in particular to the magnetic and superconducting ground states that often are not well described by the Fermi liquid model \cite{Pfleiderer2009,Brando2016,Mydosh2017}. Another line of research on uranium compounds are studies on unique magnetic transitions accompanied by a structural transition, due to strong magnetoelastic interactions. Well-known examples are insulating $\mathrm{UO}_2$ \cite{Faber1976, Giannozzi1987, Santini2009, Jaime2017} or intermetallic $\mathrm{UPd}_3$ \cite{Andres1978, McEwen1993, Walker2006}. 
	
	More recently, it has been demonstrated that $\mathrm{U}_2\mathrm{Rh}_3\mathrm{Si}_5$ also shows such a strong coupling between magnetic order and the lattice degrees of freedom. Various experiments suggest that it is a rare example of a 5$f$ material with a first order antiferromagnetic phase transition \cite{Becker1997,Takeuchi1997,Feyerherm1997,Leisure2005}. Tentatively, this was explained with the so called bootstrapping-effect, in which the crystal field splitting in combination with the magnetoelastic interactions occurring close to a magnetic transition leads to changes in the crystal field scheme \cite{Leisure2005,Wang1968,Wang1969}. This in turn influences the magnetic ordering and the structural behaviour.
	
\begin{figure}[t]
		\includegraphics[scale=0.23]{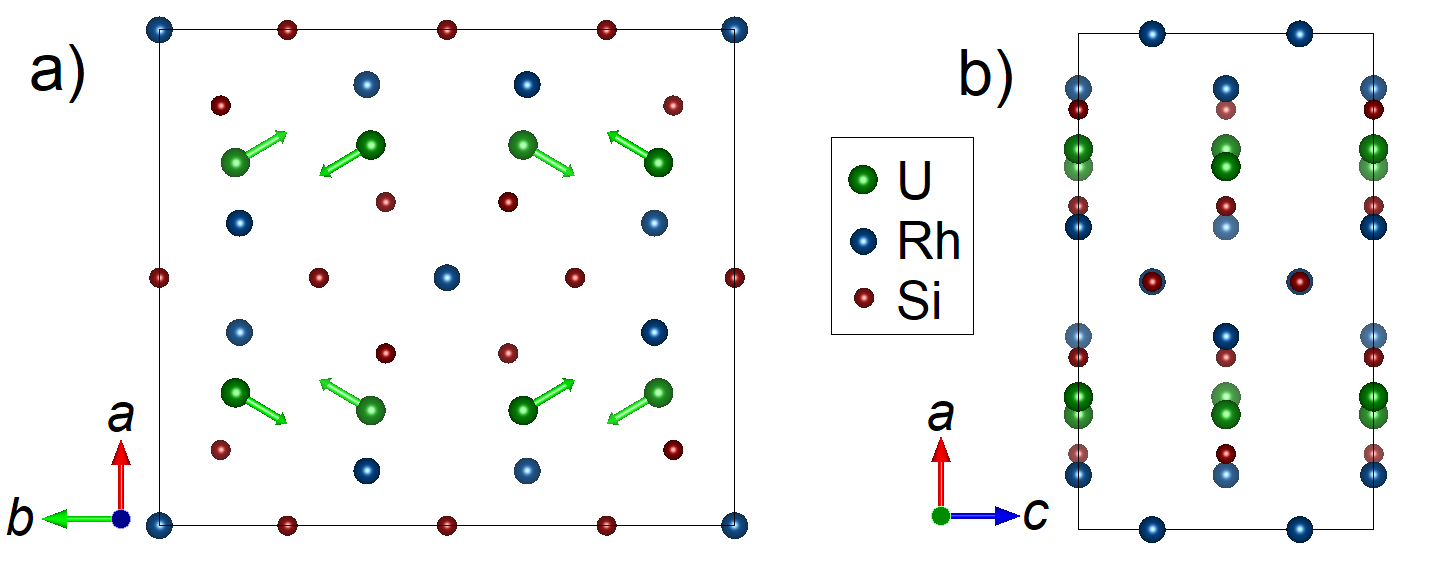}
		\caption{\label{structure} (colour online) Crystallographic and magnetic structure of $\mathrm{U}_2\mathrm{Rh}_3\mathrm{Si}_5$ from \cite{Feyerherm1997} with the view of the a) $ab$ and b) $ac$ plane; for details see text.}
	\end{figure}

	$\mathrm{U}_2\mathrm{Rh}_3\mathrm{Si}_5$ crystallizes in the monoclinic $\mathrm{Lu}_2\mathrm{Co}_3\mathrm{Si}_5$ structure with the space group $C2/c$ \cite{Hickey1990,Feyerherm1997}. The monoclinic distortion is small, with a monoclinic angle of $\beta = 90,045(10)^\circ$. Therefore, it is a common procedure to describe the crystal structure as a quasiorthorhombic lattice with the space group $Ibam$ (see Fig. \ref{structure}). The $b$ and $c$ axes are perpendicular to each other and the new direction $a'$ is specifically chosen to be perpendicular to $b$ and $c$. In the following, the direction $a'$ will be labelled the $a$ axis.
	
	$\mathrm{U}_2\mathrm{Rh}_3\mathrm{Si}_5$ orders antiferromagnetically at a temperature of $T_N~=~25.5~-~25.7~\mathrm{K}$, with the variation of $T_N$ reflecting different references, {\it i.e.}, different experimental techniques to determine this value \cite{Feyerherm1997, Becker1997, Takeuchi1997, Leisure2005}. Becker \textit{et al.} observed a sharp jump in the specific heat at $T_N$ with an amplitude of more than $100~\mathrm{J/(mol~K)}$ \cite{Becker1997}. Furthermore, a neutron-diffraction study detected a sublattice magnetization jump at $T_N$ from zero to 2/3 of its maximum within a temperature range of $0.2~\mathrm{K}$, while x-ray diffraction measurements revealed a significant expansion of the unit cell with cooling below $T_N$ \cite{Feyerherm1997}. These effects indicate a first order magnetic transition. In addition, the magnetic structure was investigated by neutron diffraction, revealing that the magnetic moments are confined to the $ab$ plane (see Fig. \ref{structure}). More specifically, they align along the direction of the nearest-neighbour U-Rh bonds \cite{Feyerherm1997}. The uranium ions have a moment of $\mu~=~2.35~\mathrm{\mu_B}$ and the linear specific-heat term of $\gamma = 22~ \mathrm{mJ/(K^2~mol)}$ suggest that they are well localized \cite{Becker1997}.
	
	Additionally, the magnetization at $4.5~\mathrm{K}$ up to 30 T was measured by Takeuchi \textit{et al.} and shows a strong magnetic anisotropy for the different crystallographic axes \cite{Takeuchi1997}. Especially the magnetization for $B||b$ axis is striking because of a sharp jump at $14~\mathrm{T}$. The large jump-like change of the magnetization by $1.6~\mathrm{\mu_B/U}$-atom supports the notion of a first order phase transition. In contrast, in these initial measurements the magnetization at $4.5$\,K for the configurations $B||a$ and $B||c$ axes increased linearly without any transitions up to highest measured fields. For all axes the magnetization appeared not to be in full saturation at $30~\mathrm{T}$. For the $b$ axis it was argued that the residual quantitative mismatch between a high-field magnetization of $1.8~\mathrm{\mu_B}$ per U-atom and an ordered magnetic moment of $2.35~\mathrm{\mu_B}$ reflects residual moment canting in the polarized state \cite{Galli2000}. For the other crystallographic directions the moments of a fraction of $1~\mathrm{\mu_B}$ suggest that there must be magnetic transitions at higher fields into the fully polarized state \cite{Takeuchi1997}.
	
	In this situation, here we present magnetization and axial magnetostriction measurements in pulsed magnetic fields up to $65~\mathrm{T}$ to extensively characterize the magnetic phase diagrams of 	$\mathrm{U}_2\mathrm{Rh}_3\mathrm{Si}_5$ for all crystallographic directions. For a comprehensive magnetic and electronic characterisation, we include the angular-dependent susceptibility, the resistivity, the magnetoresistivity and the thermal expansion in zero or moderate magnetic fields. With these data, we have been able to derive the magnetic phase diagrams of $\mathrm{U}_2\mathrm{Rh}_3\mathrm{Si}_5$ over a large temperature and magnetic field range, and have obtained new insight into the precise nature of the magnetic phase transition(s).
	
	\section{\label{sec:level2}Experimental details}
	
	For the experiment, we used three different bar shaped single crystalline samples $\mathrm{U}_2\mathrm{Rh}_3\mathrm{Si}_5$ of a few mm length and $0.5~\mathrm{x}~0.5~\mathrm{mm^2}$ cross section. The samples were cut from a  $\mathrm{U}_2\mathrm{Rh}_3\mathrm{Si}_5$ single crystal, which was grown by a modified Czochralski method in a tri-arc furnace (see: https://mgml.eu/) from a stoichiometric melt (U (3N purity), Rh (3N5 purity), Si (6N)). The uranium metal of 3N purity has been additionally purified using the Solid State Electrotransport technique \cite{Haga1998}. The pulling speed during the growth varied between 3 and 5 mm/h. The quality of the grown crystal was verified by X-ray Laue diffraction. The chemical composition of the single crystal was analysed in a scanning electron microscope Tescan Mira I LMH equipped with an Energy Dispersive X-ray (EDX) detector Bruker AXS inspecting the signal both of the secondary and backscattered electrons. Elemental mapping by EDX confirmed good composition homogeneity of the as-grown crystal. The average of multiple point scans from different parts of the sample provided the U:Rh:Si composition of 2.2(5):3.0(2):4.8(5).
	
	The axial magnetostriction and magnetization were measured in pulsed magnetic fields up to $65~\mathrm{T}$ in a temperature range from $1.4~\mathrm{K}$ to $30~\mathrm{K}$ at National High Magnetic Field Laboratory (NHMFL) at Los Alamos National Laboratory. An optical fiber with Bragg gratings was used for the measurement of the axial magnetostriction as a function of the magnetic field and the thermal expansion in zero field as a function of the temperature as described in the Refs. \cite{Daou2010,Jaime2012,Jaime2017a}. The relative values of the magnetization were measured up to $65$\,T using a pick-up coil technique \cite{Detwiler2000}. Subsequently, the magnetization was scaled onto measurements taken in a commercial SQUID magnetometer up to $5~\mathrm{T}$. In addition, the angular-dependent susceptibility has also been measured in a SQUID magnetometer with a field of $0.1~\mathrm{T}$. 
	
	Moreover, AC resistivity measurements in four-point configuration were carried out to determine the magnetoresistivity and the temperature dependence of the resistivity in magnetic fields up to $9~\mathrm{T}$ for all crystallographic axes.
	
	\section{\label{sec:level3}Results}
	
	\subsection{High field measurements}
	
	\begin{figure}[h]
		\includegraphics[scale=0.063]{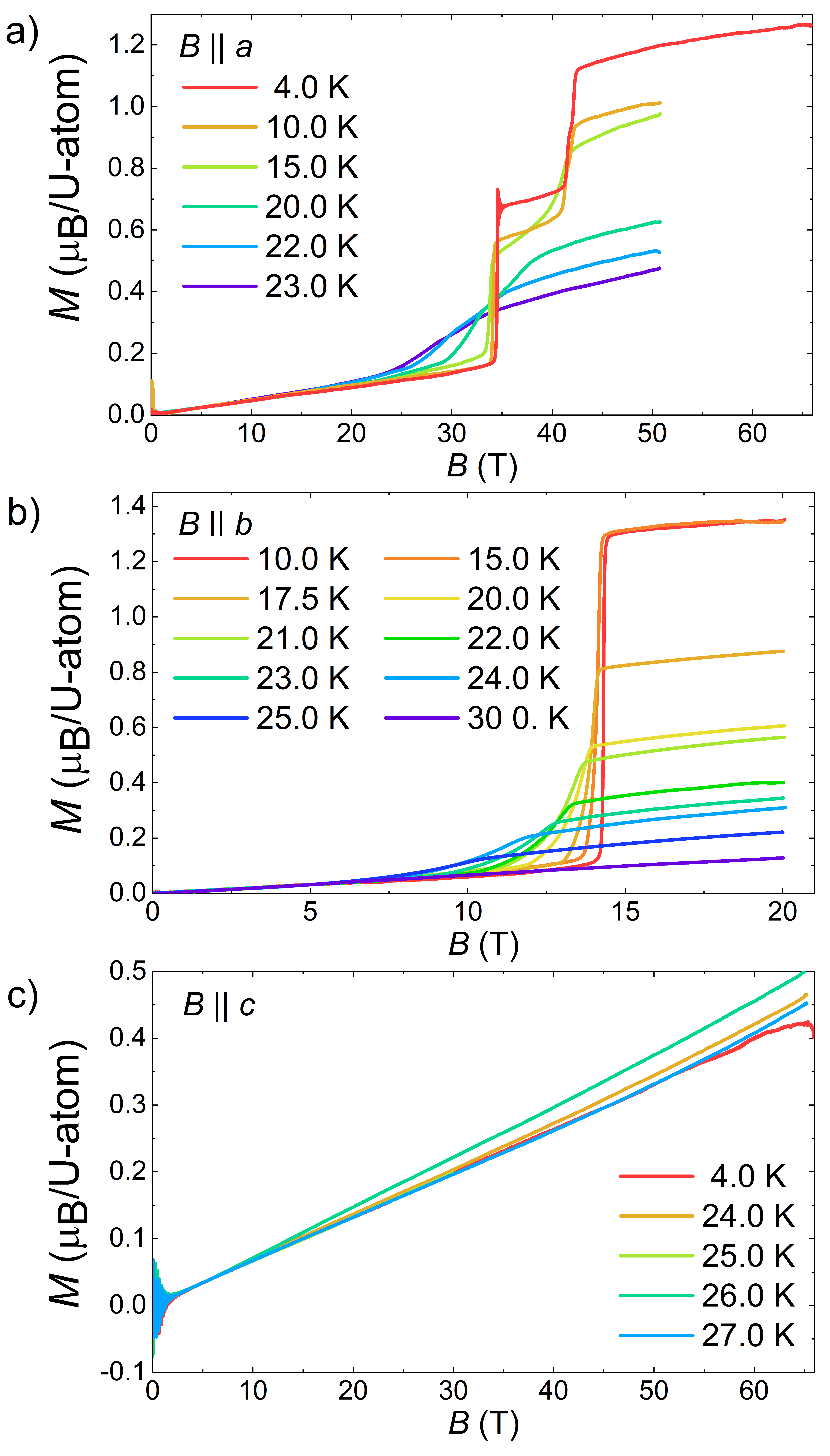}
		\caption{\label{magnetization} (colour online) Magnetization of $\mathrm{U}_2\mathrm{Rh}_3\mathrm{Si}_5$ in pulsed magnetic fields for a) $B||a$), b) $B||b$ and c) $B||c$ axes at different temperatures; for details see text.}
	\end{figure}


	The magnetization of $\mathrm{U}_2\mathrm{Rh}_3\mathrm{Si}_5$ for the different crystal axes up to $65~\mathrm{T}$ is shown in Fig. \ref{magnetization}. Overall, there is significant anisotropy to be seen in the data. Absolute values at highest fields vary between $0.5~\mathrm{\mu_B}$ ($c$ axis) and $1.4~\mathrm{\mu_B}$ per U-Atom ($b$ axis). The latter value is reasonably close to the one reported by Takeuchi \textit{et al.} \cite{Takeuchi1997} for the same field direction. Furthermore, while field induced transitions are observed for $B||a$ and $||b$ axes, this is not the case along the $c$ axis. The different features will now be discussed in detail.
	
\begin{figure}[t]
		\includegraphics[scale=0.26]{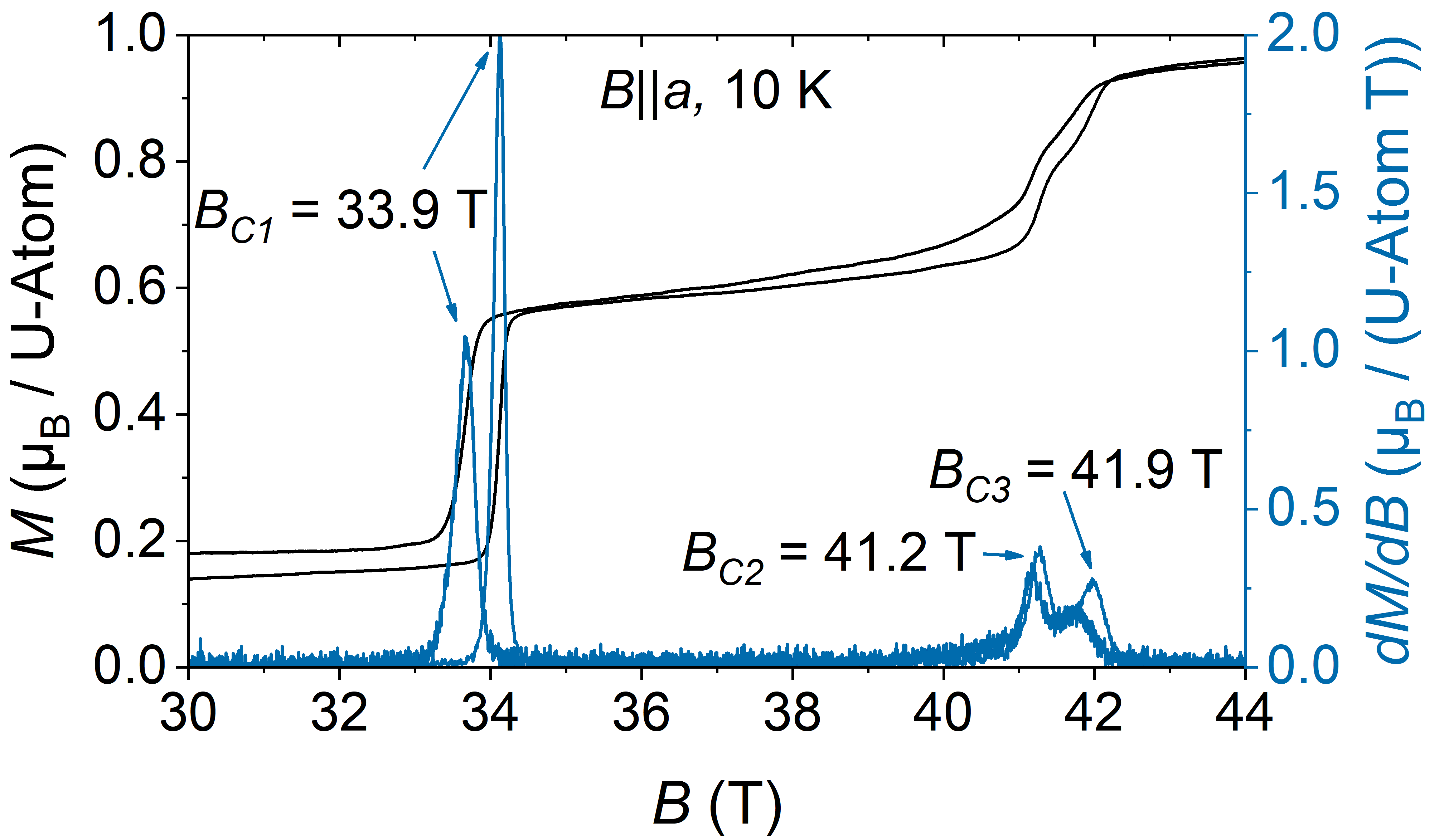}
		\caption{\label{dm_magnetization} (Colour online) High-field range of the magnetization and $\mathrm{d}M/\mathrm{d}B$ for $B||a$ axis at $10~\mathrm{K}$. The critical fields were obtained by averaging over the values of the field-up- and field-down-sweep; for details see text.}
	\end{figure}

	For the field $B||a$ axis we observe two jumps in the magnetization. At lowest temperatures, after a small linear increase the first jump appears at $34~\mathrm{T}$ from $0.1~\mathrm{\mu_B}$ up to a value of  $0.6 - 0.7~\mathrm{\mu_B}$ per U-Atom, the second occurs at approximately $42~\mathrm{T}$ with roughly the same increase of the magnetization. Closer inspection of the upper transition reveals that it appears as a two-step process by itself with two transition fields about $1~\mathrm{T}$ apart (see Fig. \ref{dm_magnetization}). It might be argued that this two-step feature is extrinsic, arising for instance from a twinned crystal with two slightly different critical fields. In that case, however, also the lower transition at $B_{C1}$ should be split into two, which is not the case and suggests the feature to be intrinsic.
	
	Above the upper transitions, for higher fields the magnetization increases again linearly with the field and is of similar magnitude as the $b$ axis magnetization. Following the argument of Galli \textit{et al.} \cite{Galli2000}, it would imply that also for the $a$ axis in high fields the magnetic moments are still canted with respect to the external field. The value of the maximal magnetization decreases as temperature is increasing and the jumps broaden significantly. Again, the sharp magnetization jumps indicate first order transitions. The transition fields used to construct the magnetic phase diagram (see below) were determined by averaging the critical field values of the field-up- and field-down-sweep. These were obtained from local maxima in $\mathrm{d}M/\mathrm{d}B$ (see Fig. \ref{dm_magnetization}).
		
	For the configuration $B||b$ axis and lowest temperatures, only one jump by $1.3~\mathrm{\mu_B}$ per U-Atom at a field of $14~\mathrm{T}$ is observed. In high fields, the maximal magnetization hardly changes for increasing temperature up to $15~\mathrm{K}$, but then decreases strongly. Additionally, the transition shifts to lower fields and broadens significantly. Again, there is an intrinsic hysteresis between the field-up- and field-down-sweep. The transition field is also determined as maximum in $\mathrm{d}M/\mathrm{d}B$.
	
	In the measurement of the magnetization for the $B||c$ axis there is no phase transition visible. The magnetization increases almost linearly with the magnetic field up to $0.5~\mathrm{\mu_B}$ per U-Atom at $65~\mathrm{T}$ at lowest temperatures and there are hardly differences for measurements at different temperatures. Moreover, it appears as if the magnetization has a small upwards curvature. It might be considered a precursor behaviour for a magnetic transition to occur at even higher fields.


	\begin{figure}[t]
		\includegraphics[scale=0.27]{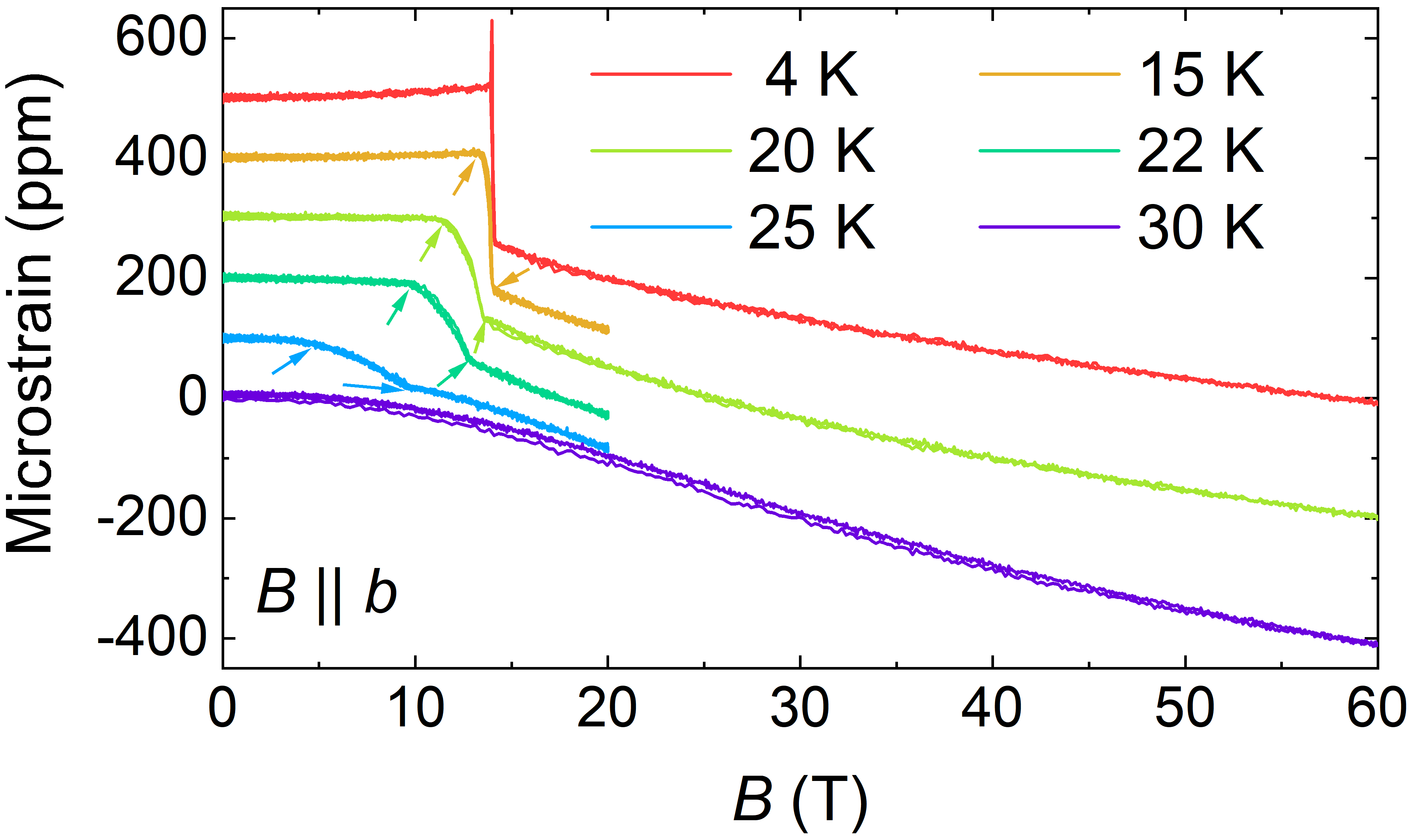}
		\caption{\label{magnetostriction} (Colour online) Axial magnetostriction of $\mathrm{U}_2\mathrm{Rh}_3\mathrm{Si}_5$ for magnetic fields along the $b$ axis for different temperatures up to $60$\,T (for clarity data are shifted with respect to each other by $100~\mathrm{ppm}$); for details see text.}
	\end{figure}

	\begin{figure}[t]
		\includegraphics[scale=0.27]{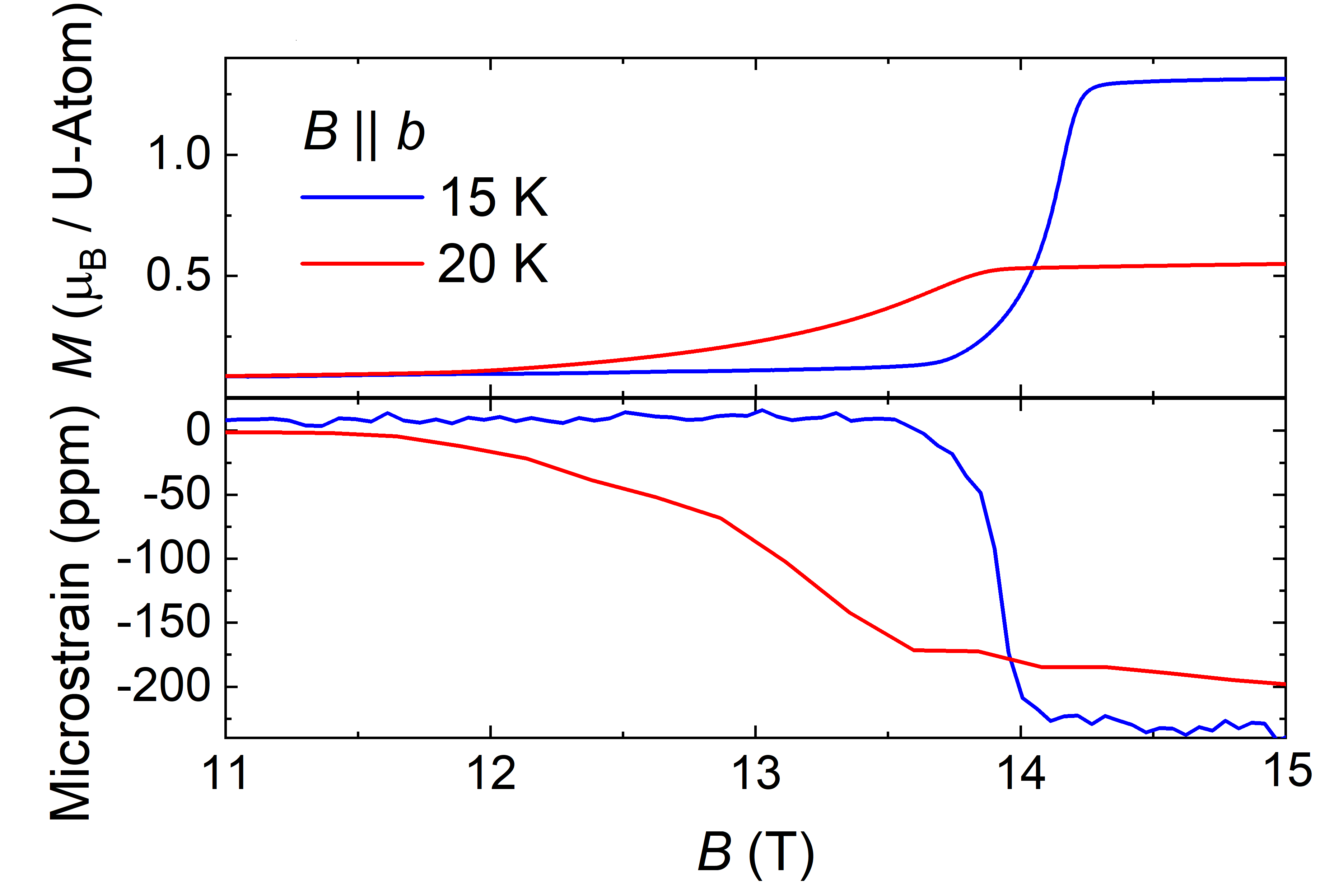}
		\caption{\label{vgl_magn_magstr} (Colour online) Comparison of the magnetization and the magnetostriction at $15$\,K and $20$\,K for $B||b$ axis; for details see text.}
	\end{figure}
	
	For additional information about the interdependence between the structure and the magnetic behaviour, we investigated the axial magnetostriction for $B||b$ axis in fields up to $60~\mathrm{T}$. The corresponding measurements along the other crystallographic axes failed because of the strong magnetic anisotropy of the samples. For field directions away from the easy magnetic axis, it leads to magnetic torque on the sample, significantly twisting it on the fiber used for the magnetostriction experiment and thus affecting the experiment.
	
	The measurement for the $B||b$ axis at different temperatures is shown in Fig. \ref{magnetostriction} (for clarity data are shifted with respect to each other by $100~\mathrm{ppm}$). At low temperatures ($4~\mathrm{K}$) the transition observed in the magnetization is clearly identified by a drop of the microstrain by $250~\mathrm{ppm}$ at $14~\mathrm{T}$. As temperature is raised, the magnitude of the drop decreases, the transition becomes wider and seems to transform into a two-step transitions. The arrows in Fig. \ref{magnetostriction} were determined as the points, where a local extrema is detectable in the second field derivative of the microstrain. This broadening of the transition is also visible in the magnetization (as shown in Fig. \ref{vgl_magn_magstr}), where with raising temperatures the transition changes from the low-temperature step-like behavior into a more S-shape form. The field value of the "high-field side" of the magnetostrictive transition is in a very good agreement with the turning point in the magnetization along the $b$ axis, while the feature at the "low-field side" of the magnetostricitive transition appears to match the beginning of the upturn in the magnetization. Because the transitions in the magnetization along the $a$ axis exhibits a similar broadening with temperature, we suspect a similar behavior of the magnetostriction for $B||a$ as for $||b$ axis.

	\subsection{Resistivity}
	
\begin{figure}[t]
		\includegraphics[scale=0.185]{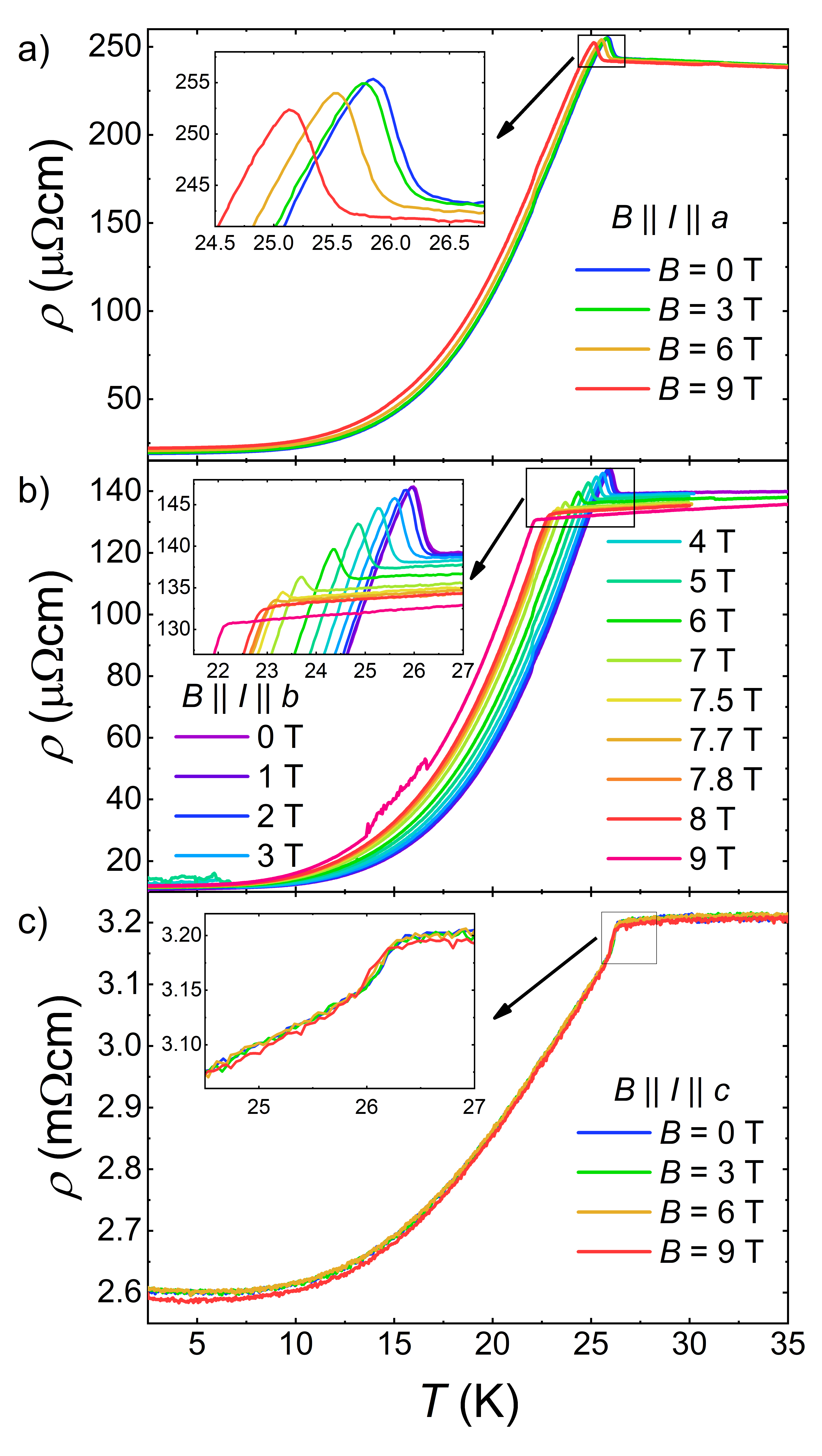}
		\caption{\label{resistivity} (Colour online) Resistivity of $\mathrm{U}_2\mathrm{Rh}_3\mathrm{Si}_5$ as a function of temperature for $a$, $b$ and $c$ axis in different magnetic fields up to $9~\mathrm{T}$; for details see text.}
	\end{figure}

	To link our high field data on the magnetic phase diagram with the low field behaviour of $\mathrm{U}_2\mathrm{Rh}_3\mathrm{Si}_5$, in a next step we measured the resistivity as a function of the temperature for different magnetic fields along the three axes (Fig. \ref{resistivity}). The residual resistivity $\rho_0$ of approximately $20~\mathrm{\mu \Omega cm}$ for the $a$ axis and $10~\mathrm{\mu \Omega cm}$ for the $b$ axis are similar to those reported in Ref.~\cite{Becker1997} and indicate a good sample quality. In contrast, the sample along the $c$ axis shows a much higher residual resistivity of $2.6~\mathrm{m \Omega cm}$. Previously, much smaller values have been reported for this axis \cite{Becker1997}. As our $c$ axis crystal stems from the same batch as the other two samples, we believe that this particular sample is microcracked, possibly as result of being cycled multiple times through the first order phase transition during the measurements, this way likely affecting the absolute value of the resistivity for this crystallographic direction. This is supported by the fact that the residual resistivity $\rho_0$ for the $c$ axis was lower in a first measurement.
	
	There are various peculiarities in the resistivity visible. In the measurements along the $a$ and $b$ axes an anomaly around the transition temperature is noticeable. In detail (see inset), in zero field the resistivity increases with decreasing temperature bellow $T^*~=~26.4~\mathrm{K}$ for the $a$ axis and $T^*~=~26.5~\mathrm{K}$ for the $b$ axis measurement, \textit{i.e.}, slightly above the antiferromagnetic (AFM) transition temperature. Here, the critical temperature $T^*$ of the upturn was determined as the maximum in the second temperature derivative of the resistivity. Only after cooling by $0.5~\mathrm{K}$, \textit{i.e.}, down to $T_N$, the resistivity turns over and decreases steeply. The maximum is found at a temperature of $~25.9~\mathrm{K}$ for the $a$ axis and $26.0~\mathrm{K}$ for the $b$ axis measurement.
	
	In an external magnetic field along the $a$ axis the anomaly slightly shifts to lower temperatures (see inset of Fig. \ref{resistivity}(a)). For the configuration $B~||~I~||~b$ the anomaly is changing much more rapidly with an increasing magnetic field (inset Fig. \ref{resistivity}(b)). First, a shift to lower temperatures with magnetic field is visible. In addition, the width and the height of the peak like anomaly is decreasing with increasing field. At a field of $7.8~\mathrm{T}$ the anomaly is not visible any more and the resistivity immediately decreases at the transition temperature. In contrast, there is no peak like anomaly measured for the $c$ axis. The resistivity instead exhibits a relatively sharp downturn at the temperature, where the upturn in the resistivity is visible for the $a$ and $b$ axes, and shows a kink at $T_N$. External magnetic fields up to $9~\mathrm{T}$ along the $c$ axis have no impact on these features.
	
	Previously, Becker \textit{et al.} \cite{Becker1997,Becker1999} measured the resistivity in zero field for the three axes and in fields up to $20~\mathrm{T}$ for the $b$ axis. In their interpretation, the peak like anomaly was associated to the opening of a superzone gap due to antiferromagnetic ordering, as it is observed in Erbium and Terbium \cite{Mackintosh1962}. Such a gap leads to a rounded maximum of the resistivity just below the transition temperature as result of a modified Brillouin zone in the magnetically ordered phase. The drop of the resistivity upon further lowering temperature then is due to the reduction of phonon and magnon scattering. This interpretation seems to match the fact, that the anomaly is only visible for the $a$ and $b$ axis. The magnetic moments lie in the $ab$ plane and therefore the unit cell only doubles along the $a$ and $b$ directions at the antiferromagnetic transition.
	
	However, there are significant differences to superzone gap occurrences such as for Erbium and Terbium \cite{Mackintosh1962}. In U$_2$Rh$_3$Si$_5$, the upturn in the resistivity for the $a$ and $b$ axes and the downward drop for the $c$ axis take place at a temperature of $0.5~\mathrm{K}$ above the transition temperature $T_N$. Moreover, at $T_N$, the resistivity of our U compound for the $a$ and $b$ axes shows a kink-like downturn, while for the $c$ axis we observe a change of slope. This is in contrast to the rounded maxima seen in Erbium and Terbium and predicted by theory \cite{Elliott1963}. In addition, an external magnetic field does not lead to a gradual disappearance of the gap (see other materials with a superzone gap like Terbium \cite{Hegland1963} or $\mathrm{URu_2Si_2}$ \cite{Mentink1996}). After all, the resistive feature associated to the occurrence of a superzone gap is related to the change of the translationally invariant cell in the antiferromagnetic phase, which is not a continuous function of the magnetic field. In conclusion, the interpretation of the peak like anomaly in $\mathrm{U}_2\mathrm{Rh}_3\mathrm{Si}_5$ as result of a superzone gap may be incorrect and other effects appear to be responsible for the peculiar behaviour of the resistivity around the transition temperature. We note that, while the discrepancy between the upturn temperature $T^*$ and the antiferromagnetic transition temperature $T_N$ in U$_2$Rh$_3$Si$_5$ was not reported before, close inspection of the plots in Ref. \cite{Becker1997} suggest a similar discrepancy to exist for those data.

	
	\begin{figure}[t]
		\includegraphics[scale=0.2]{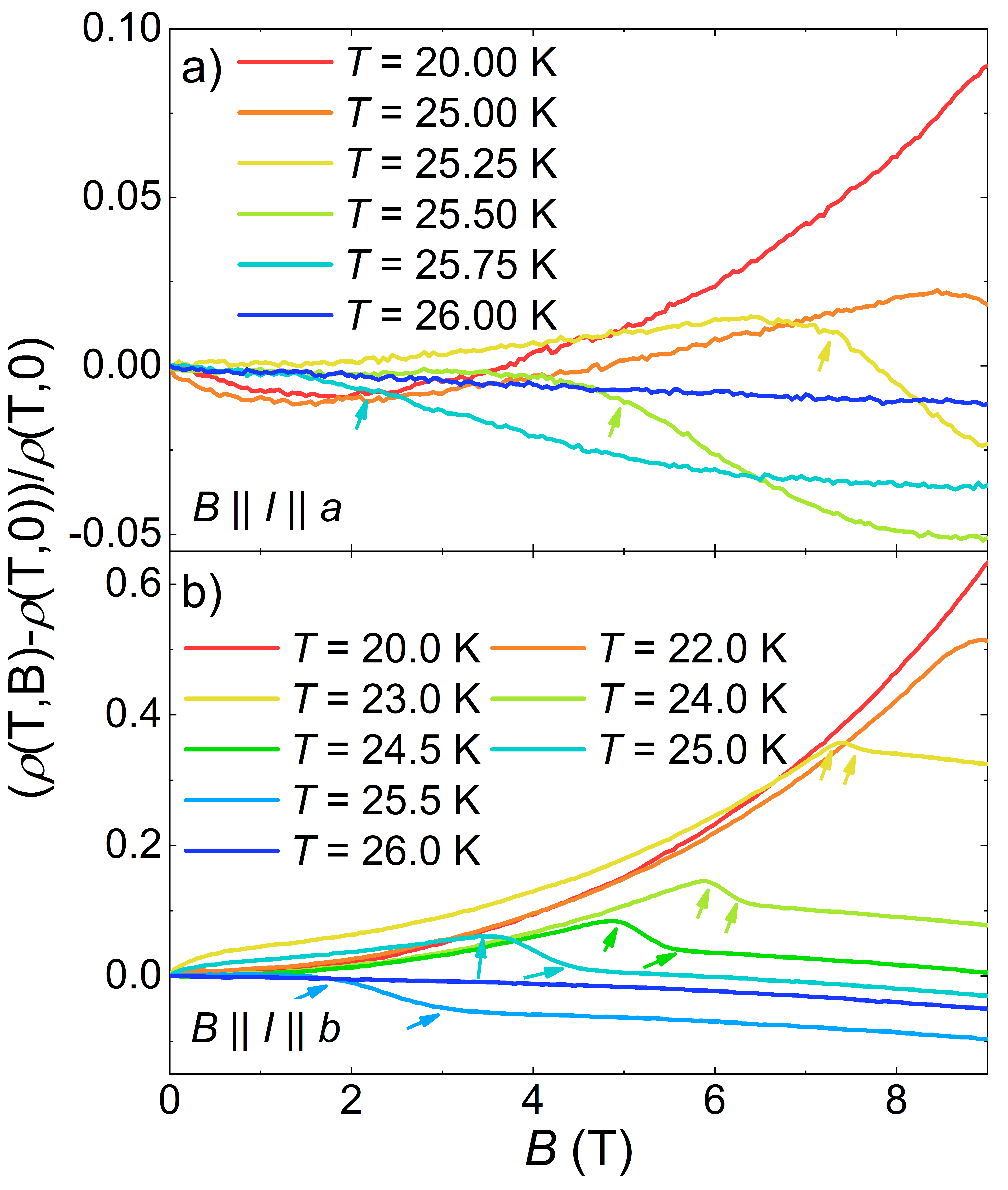}
		\caption{\label{magnetoresistance} (Colour online) Longitudinal magnetoresistivity of $\mathrm{U}_2\mathrm{Rh}_3\mathrm{Si}_5$ for magnetic fields along the a) $a$ axis and b) $b$ axis for different temperatures up to $9~\mathrm{T}$; for details see text.}
	\end{figure}
	
	Finally, the longitudinal magnetoresistivity (MR) along the three crystallographic axes was measured. In Figure \ref{magnetoresistance} we plot the normalized magnetoresistivity as function of the external magnetic field for different temperatures. We omit the magnetoresistivity measured for $B~||~I~||~ c$ axis, as there was no resolvable signal beyond experimental scatter, this likely as result of the large extrinsic residual resistivity for this axis.

	The magnetoresistivity is highly anisotropic. For instance, at $20~\mathrm{K}$ we find a MR (defined as ${(\rho(T,B)-\rho(T,0))/\rho(T,0)}$) of up to about 10$~$\% along the $a$ axis, while along $b$ it reaches 60$~$\%. This obviously reflects the much stronger field dependence of $T_N$ along the $b$ axis than along $a$ and is consistent with the temperature-dependent resistivity measurements (Fig. \ref{resistivity}), where a magnetic field caused large changes of the resistivity for the $b$ axis and only moderate changes for the $a$ axis.

	Still, the measurements for the configurations $B~||~I~||~a$ and $||~b$ show qualitatively a similar behaviour. In detail, discussing the $b$ axis data, starting from high fields and for temperatures in a range from $23~\mathrm{K}$ to $26~\mathrm{K}$, the magnetoresistivity rises monotonously upon lowering the field down to the upper critical field. At the critical field there is a two step transition visible (indicated by arrows). First the slope of the magnetoresistivity changes and rises more steeply until a local maximum develops. This two step transition corresponds to the feature in the temperature dependent resistivity (Fig. \ref{resistivity}). The upturn in the resistivity shows up as a change in the slope of the magnetoresistivity and the maximum in the resistivity is visible as a maximum in the magnetoresistivity. After this maximum the magnetoresistivity falls off monotonously. For the $a$ axis data, only the local maximum in the MR is clearly visible, while the change of slope is not easily identified. This however might simply reflect the stretched field scale for the $a$ axis compared to $b$, making it harder to identify the upper critical field in the experimental window we access.


	\subsection{Susceptibility}
	
	\begin{figure}[t]
		\includegraphics[scale=0.24]{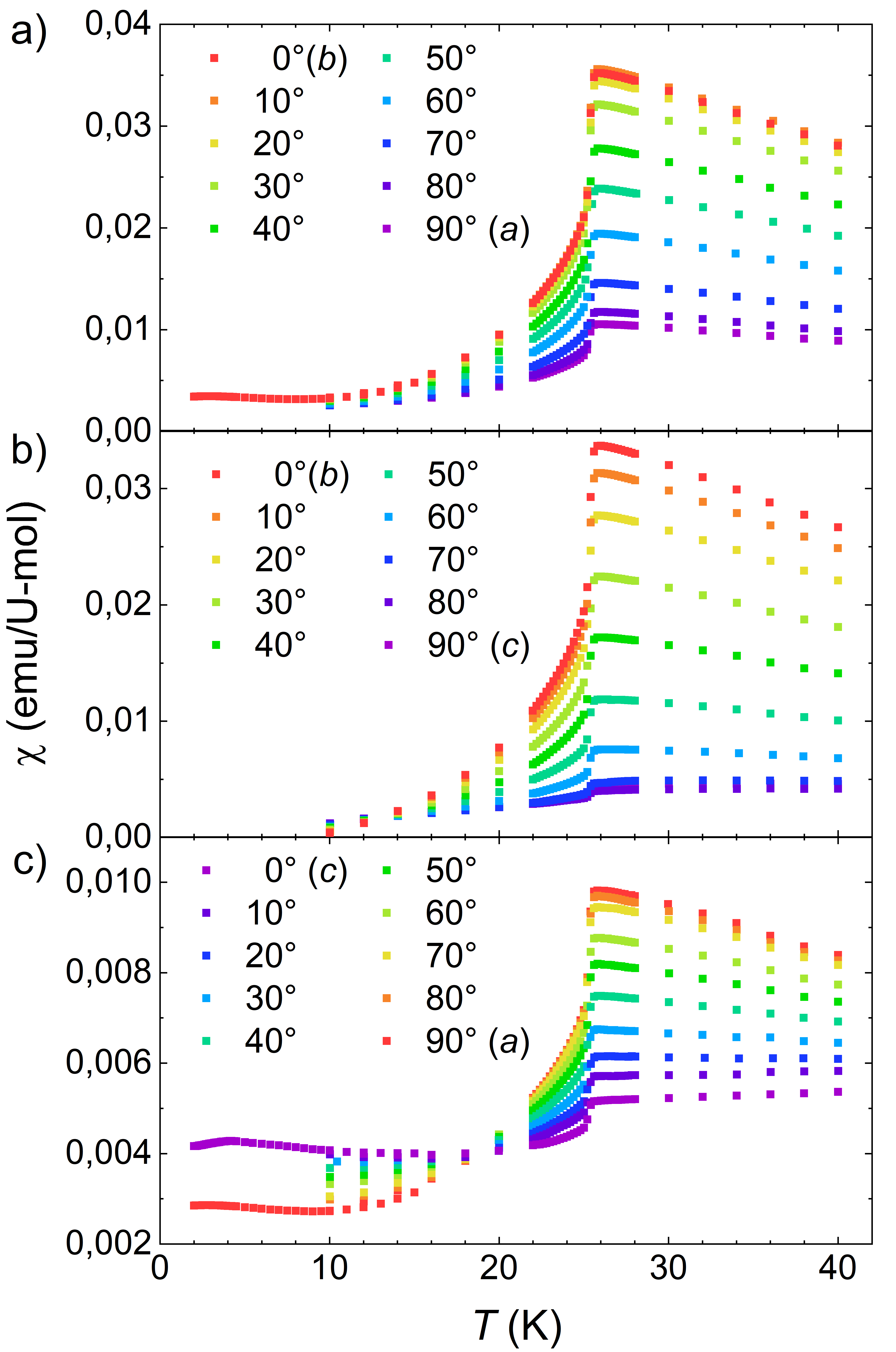}
		\caption{\label{susceptibility} (Colour online) Angular-dependent measurements of the susceptibility with $10^\circ$ steps for rotations from a) $b$ to $a$ axis, b) $b$ to $c$ axis and c) $c$ to $a$ axis; for details see text.}
	\end{figure}

	Since the magnetization and magnetostriction experiments revealed a very large magnetic anisotropy, we measured the angular dependent susceptibility in small magnetic fields ($0.1~\mathrm{T}$). Fig. \ref{susceptibility} shows the susceptibility of U$_2$Rh$_3$Si$_5$ in emu/U-mol as it is rotated from the $b$ axis in $10^\circ$ steps towards the $a$ and the $c$ axis, and for the corresponding rotation from the $c$ towards the $a$ axis.
	
	For high temperatures the susceptibility shows a Curie like behaviour down to $T_N~=~25.8~\mathrm{K}$, where the susceptibility drops because of the antiferromagnetic transition. The almost steplike behaviour at the transition supports the assumption of a first order transition. As noted before, the anisotropy between the three axes is very large. In the paramagnetic phase the biggest susceptibility response is detectable for the $b$ axis with a maximum of the susceptibility at $T_N$ of $0.035~\mathrm{emu/U{-mol}}$, while for the $a$ ($c$) axis the maximum susceptibility is a factor of 3 (7) smaller. Therefore, we conclude that the $b$ axis is the magnetically easiest and the $c$ axis the hard axis. The anisotropy in the magnetically ordered phase is consistent with this view and the reported magnetic structure, as the magnetic moments are oriented within the $ab$ plane with an angle of $34^\circ$ to the $b$ axis.

	\begin{figure}[t]
		\includegraphics[scale=0.095]{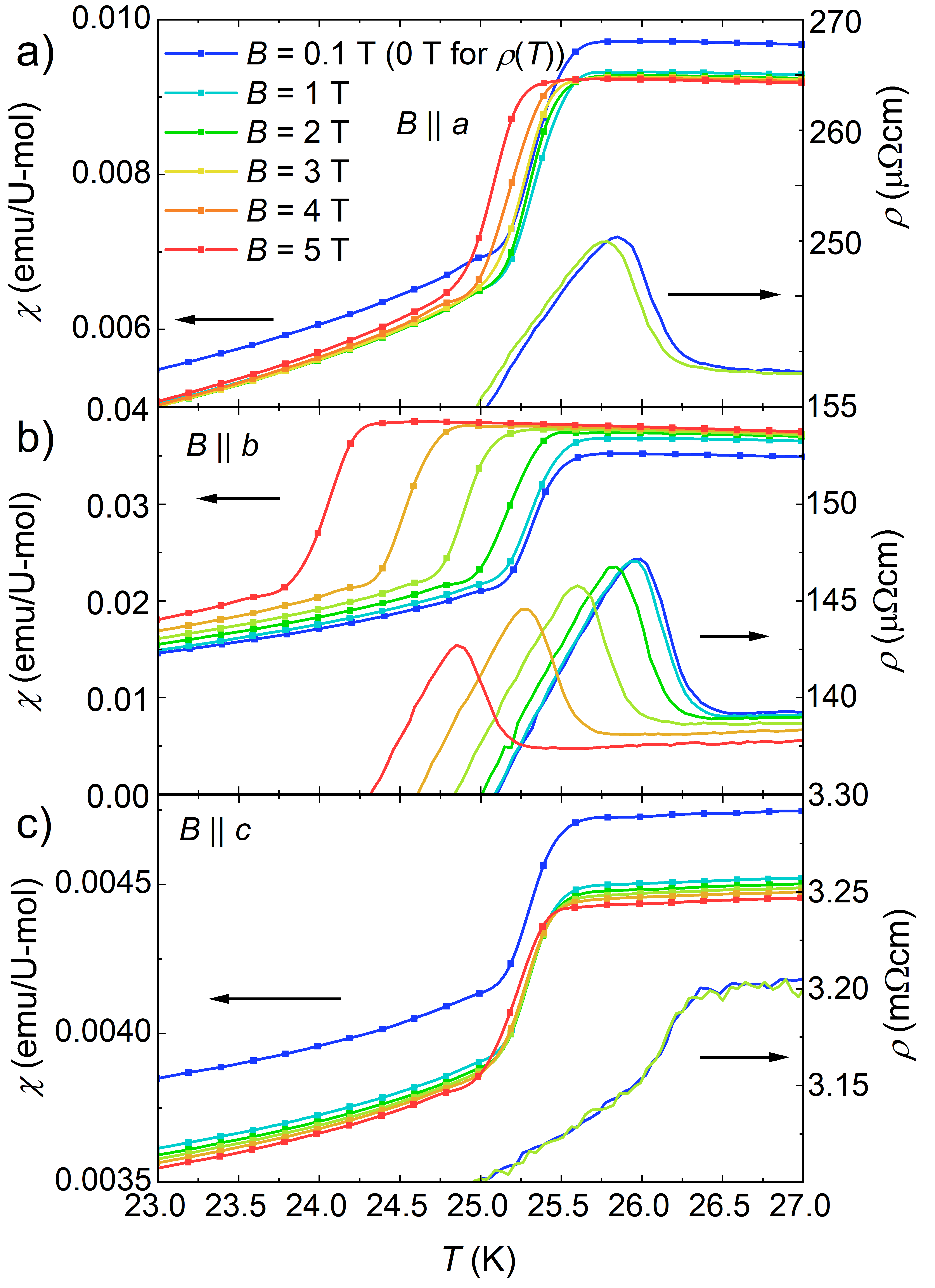}
		\caption{\label{susceptibility_T} (Colour online) Measurements of the susceptibility for a) $B||a$, b) $B||b$ and c) $B||c$ between $23$\,K and $27$\,K for fields up to $5$\,T. For direct comparison the corresponding resistivity measurements already shown in Fig. \ref{resistivity} are included; for details see text.}
	\end{figure}

	To complete our investigation of the phase transitions, we measured the temperature dependent susceptibility in fields up to $5$\,T and compare it with the corresponding resistivity data in Fig. \ref{susceptibility_T}. Beginning with the $a$ axis, the susceptibility transition shifts only moderately to lower temperatures from $T_N~=~25.8$\,K at $0.1$\,T to $T_N~=~25.6$\,K at $5$\,T. The transition temperature was determined as the onset of the change of slope of the susceptibility upon lowering temperatures from above $T_N$. The absolute values of the susceptibility differ only slightly for the measurements in different fields, with a tendency to an increase of the height of the step-like transition at $T_N$. An exception is the measurement at $0.1$\,T, which is about $0.5~\mathrm{memu/U}$-mol larger than those at higher fields. It might reflects a residual contribution from a small amount of paramagnetic impurities, which are saturated in higher fields. 
	
	In comparison, the influence of the magnetic field on the $b$ axis susceptibility is stronger. While $\chi (T)$ decreases approximately by $0.014~\mathrm{emu/U}$-mol in $0.1$\,T at the transition from paramagnetism into the antiferromagnetic phase, it decreases by $0.018~\mathrm{emu/U}$-mol in $5$\,T with a transition temperature of $T_N~=~24.5$\,K. Thus, the AFM transition in the susceptibility shifts to lower temperatures and is more pronounced for higher fields. Finally, the susceptibility for the $c$ axis shows basically no field dependence up to $5$\,T. Only, similar to the $a$ axis, the susceptibility at $0.1$\,T is a bit higher possibly because of paramagnetic impurities.  
	
	To illustrate the field dependence of the antiferromagnetic transition seen in the susceptibility and resistivity, in Fig. \ref{susceptibility_T} we include the latter quantity. The comparison shows that the transition temperature in the susceptibility for the $a$ and $b$ axes are in reasonable good agreement \cite{newDis} (difference of $0.1$ to $0.2$\,K) with the maxima in the resistivity measurements, and not with the upturn in $\rho$. Therefore, it supports our conclusion that the upturn in the resistivity is not due to the antiferromagnetic transition. Analogously, the transition temperature of the $c$ axis in the susceptibility fits better the second change of the slope in the resistivity, when coming from higher temperatures.
	
	\subsection{Thermal expansion}
	
	\begin{figure}[t]
		\includegraphics[scale=0.31]{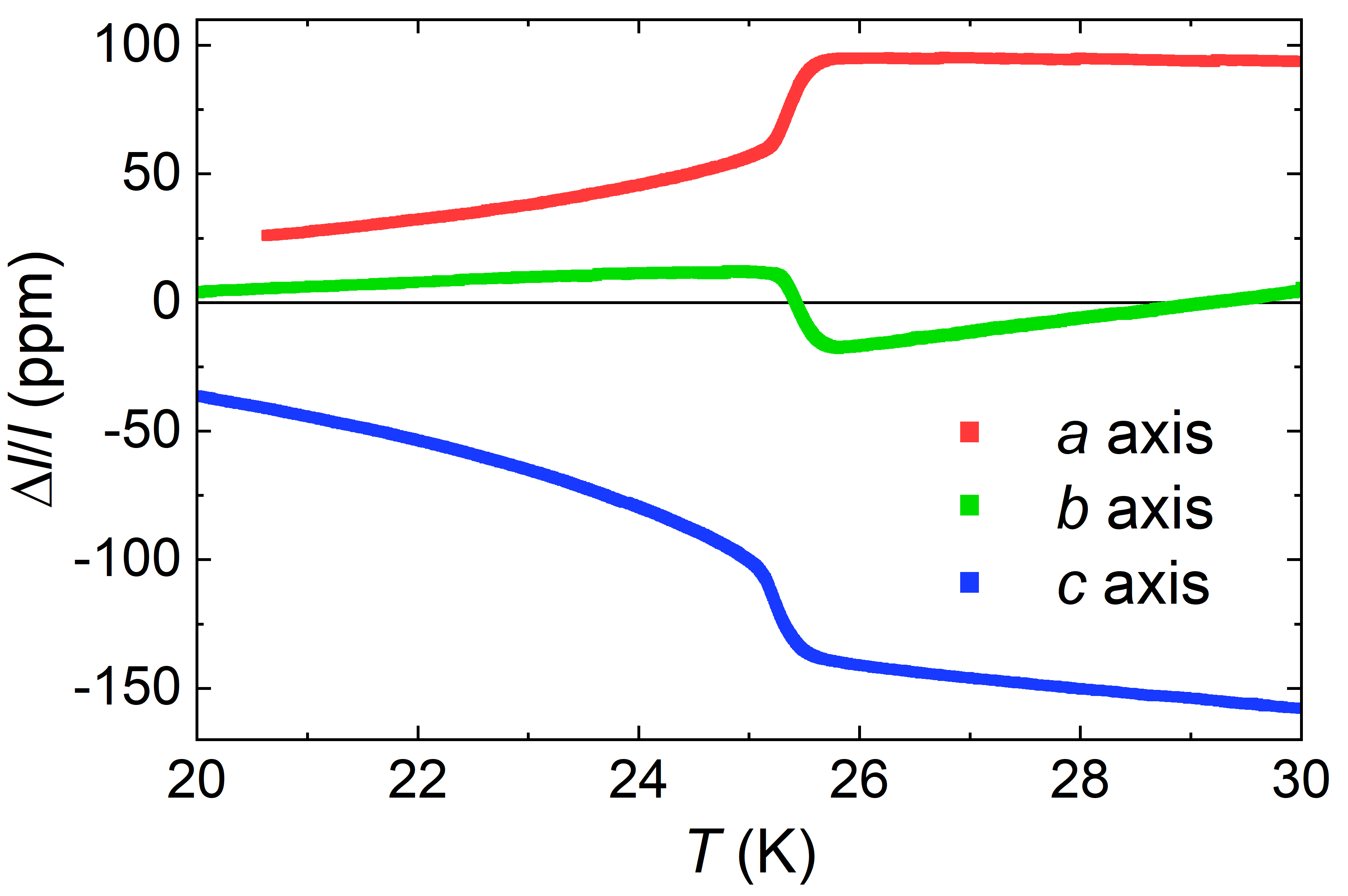}
		\caption{\label{expansion} (Colour online) Comparison of the magnetization and the magnetostriction at $15$\,K and $20$\,K for $B||b$; for details see text.}
	\end{figure}
	
	In order to affirm the results from previous x-ray diffraction experiments \cite{Feyerherm1997}, we measured the thermal expansion in zero magnetic field. In Fig. \ref{expansion} we present the thermal expansion $\Delta l/l$ in ppm as function of temperature from $20$\,K to $30$\,K for the three axes. With cooling down the unit cell of $\mathrm{U}_2\mathrm{Rh}_3\mathrm{Si}_5$ expands slightly in the $a$ and $c$ direction and contracts along the $b$ axis. At the antiferromagnetic transition temperature of $25.8$\,K the thermal expansion changes drastically along all axes, leading to an expansion along the $b$ and $c$ axes and a contraction along the $a$ axis. The change of the thermal expansion is approximately $\delta_a=38$\,ppm, $\delta_b=-29$\,ppm and $\delta_c=-39$\,ppm within $0.8$\,K of $T_N$. After this jump in $\Delta l/l$ the unit cells further expands along the $c$ axis and contracts along the $a$ and $b$ axes.
	
	Our thermal expansion measurement is in very good agreement with the measurement of the lattice parameters by x-ray diffraction \cite{Feyerherm1997}. The pronounced jump in the thermal expansion is triggered by the antiferromagnetic first order transition. Notably, the transition temperature is similar to the susceptibility at $25.8$\,K \cite{newDis}. Thus, the lattice response in $\mathrm{U}_2\mathrm{Rh}_3\mathrm{Si}_5$ is directly impacted by the magnetism. This supports our assumption that the maximum in the resistivity at $25.8$\,K is due to the magnetic first order transition and the upturn in the resistivity for the $a$/$b$ axis at $26.4$\,K/$26.5$\,K is caused by something different.
	

	\subsection{Magnetic phase diagram}
	
\begin{figure}[t]
		\includegraphics[scale=0.12]{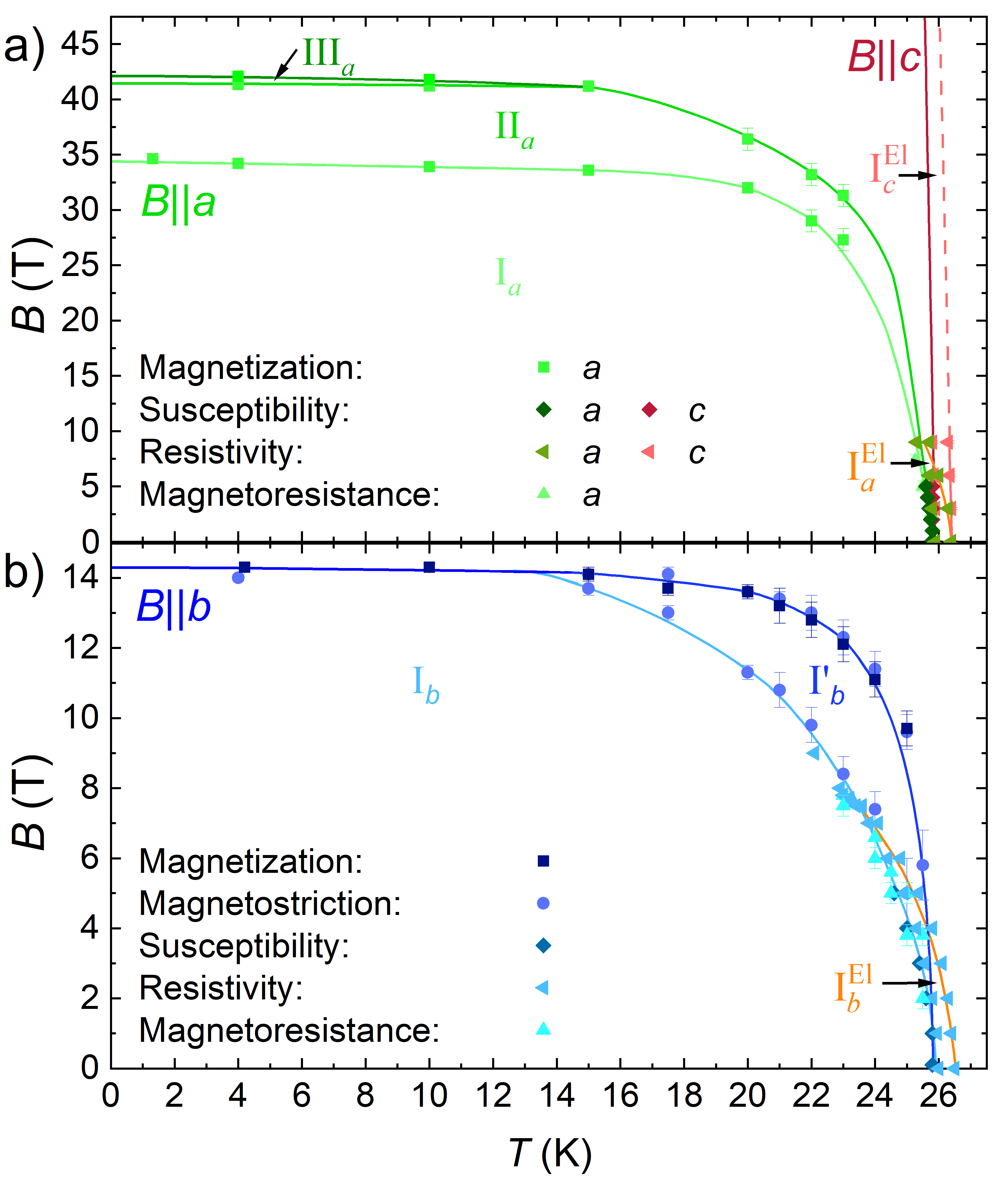}
		\caption{\label{diagram} (Colour online) Phase diagram of $\mathrm{U}_2\mathrm{Rh}_3\mathrm{Si}_5$ for a) the $a$ and $c$ and b) for the $b$ axis over a wide temperature and magnetic field range. The solid lines are guides to the eye denoting phase border lines, while the dashed lines are suspected  phase border lines; for details see text.}
	\end{figure}

	From our data we construct the magnetic phase diagram of $\mathrm{U}_2\mathrm{Rh}_3\mathrm{Si}_5$ for magnetic fields up to $65~\mathrm{T}$ in Fig. \ref{diagram}. The data points have been collected from magnetization, magnetostriction, magnetic susceptibility, resistivity and magnetoresistivity as described above. We start by discussing the magnetically ordered phases.
	
	In zero field, AFM order sets in below $T_N~=~25.8~\mathrm{K}$. The AFM ground state phase is labelled I$_a$ along the $a$ axis, and correspondingly I$_b$ along the $b$ axis (Fig. \ref{diagram} a)). Notably, in magnetic fields there is a clear and pronounced anisotropy visible for the different axes. The phase boundary for the AFM phase I$_a$ at low temperatures is found at $34~\mathrm{T}$, while the phase boundary for I$_b$ lies only at fields of $14~\mathrm{T}$. In contrast, there was no phase transition observed for the $c$ axis up to $65~\mathrm{T}$ in the magnetization, implying that for this axis below $T_N$ and the experimental field range the system is always in the AFM phase I. 
	
	
	For the $c$ axis data, we note that no phase boundary has been crossed even at $25$\,K in $65$\,T. If we compare this observation to the $a$ or $b$ axis data, where at $25$\,K the critical field is of the order of half of the zero temperature value, it implies that for the $c$ axis the corresponding zero temperature critical field value would be about $100$\,T or more. Thus, overall we find an anisotropy of the critical fields of $\mathrm{U}_2\mathrm{Rh}_3\mathrm{Si}_5$ of at least a factor of 5 between $b$ and $c$ axis.
		
	We will now discuss the details of these phase diagrams, starting with the magnetic transitions of the $a$ axis. There are two clear jumps visible in the magnetization measurement (with the upper of these jumps as a two-step transition at low temperatures) and therefore, we assume that these are first order transitions. The new phase II$_a$ (Fig. \ref{diagram}(a)) was measured for a broad temperature and magnetic field range. The border lines of phase I$_a$ and II$_a$ basically evolve parallel to each other with a distance of $7-8~\mathrm{T}$ up to a temperature of about 15 K. Then, for temperatures up to $T_N = 25.8~\mathrm{K}$, the critical fields rapidly drop to zero. Additionally, the two-step transition measured in the magnetization at $4$\,K and $10$\,K reflects a narrow phase III$_a$ with a width of $1$\,T prior to the field-polarized phase. Since the two-step transition is not visible in the magnetization at $15$\,K, we assume that the phase III$_a$ only exist up to a temperature of $10$ to $15$\,K.

	In comparison, for the easy axis, \textit{i.e.}, the $b$ axis (see Fig. \ref{diagram}(b)) we resolve additional peculiarities. Our measurements reveal a first order phase transition at $14~\mathrm{T}$ for low temperatures. At the phase transition the magnetic moments flip from the AFM phase I$_b$ into a field-polarized phase. Surprisingly, in the magnetostriction the step-like drop observed at lowest temperatures transforms into a double transition (Fig. \ref{magnetostriction}) in a temperature range from $15~\mathrm{K}$ to $25.8~\mathrm{K}$, denoting a distinct phase range I'$_b$ (Fig. \ref{diagram}(b)). This change of the field-dependent behaviour is also reflected in the magnetization, which in this temperature range exhibits the more gradual character of a metamagnetic transition, which starts at the borderline I$_b~\rightarrow~$I'$_b$.  At the upper boundary of phase range I'$_b$, visible in the magnetization as magnetic saturation and in the magnetostriction as a gradually flattening behaviour, the magnetic moments flip into the field polarized state.
	
	The magnetization of the $a$ axis shows a similar gradual character of the metamagnetic transition in a similar temperature range. Therefore, we assume, that the $a$ axis exhibits a phase range I'$_a$ equivalent to I'$_b$ for the $b$ axis. As we will discuss below, in particular resistivity measurements in fields up to $\sim 25$\,T should be a suitable tool to identify this phase range.
	 
	Finally, we address the $c$ axis data, where we have only been able to find a transition in the temperature dependent susceptibility and resistivity, but not in the field dependent magnetization. From our non-observation of a phase transition in our pulsed field experiments we conclude that the magnetic fields have not been strong enough to align the magnetic moments along the $c$ axis. This thought is supported by the alignment of the magnetic moments in the $ab$ plane in zero field reported in Ref. \cite{Feyerherm1997}. Therefore, the phase boundary for the $c$ axis is very steep as indicated by the red line in Fig. \ref{diagram}(a).
	
	Aside from the transitions detected in thermodynamic or structural properties, in addition very unusual features are visible in the resistivity. As pointed out, the upturn of the resistivity with decreasing temperature for the $a$ and $b$ axis and the downward jump for the $c$ axis are at slightly higher temperatures $T^*$ than $T_N$. Moreover, these anomalies appear not to be visible in the susceptibility, thermal expansion and magnetization measurements. We plot the field evolution of this transition with an orange (red) line for the $a$ ($c$) axis in Fig. \ref{diagram}(a) and with an orange line for the $b$ axis in Fig. \ref{diagram}(b). A close look at the $b$ axis indicates that for zero/low magnetic fields this phase border line is clearly distinct from both the phase boundaries I$_b/$I'$_b$ and I'$_b/\mathrm{PM}$. However, with increasing field this transition merges into the lower phase border line I$_b/$I'$_b$. The merger denotes the point where no upturn in the temperature dependent resistivity is seen any more, that is at 7.8\,T. Thus, by our data, we conclude that for the $b$ axis experiment there is even a distinct third phase in the phase diagram of $\mathrm{U_2Rh_3Si_5}$, \textit{i.e.}, phase I$^{\mathrm{El}}_b$. As it is only detected in the resistivity, it appears to be an electronic rather than a magnetic transition.
	
	Given that along the $a$ axis the overall resistive behavior is very similar to the $b$ axis, it is likely that also for this axis in higher fields the upturn in the resistivity disappears. The behavior would then be analogous to the $b$ axis, implying that also for this crystallographic direction there appears a distinct phase I$^{\mathrm{El}}_a$ (see Fig. \ref{diagram}(a)). Here, additional resistivity measurements in higher fields up to the $20$\,T range are called for. Finally, while for the $c$ axis the experimental data is less abundant, the temperature range of the almost step-like reduction of the resistivity similarly would signal a distinct electronic phase I$^{\mathrm{El}}_c$ (Fig. \ref{diagram}(a)) for this crystallographic direction.

	\section{\label{sec:level4}Discussion}
	
	We will now summarize our experimental findings and discuss the implications of the new observations. Overall, from the broad characterization of our single crystalline samples U$_2$Rh$_3$Si$_5$, their physical properties correspond to those previously reported. In particular, our measurements fully agree with the notion of a first order nature of the antiferromagnetic transition at $T_N$ \cite{Becker1997,Takeuchi1997,Feyerherm1997,Leisure2005}.
	
	Notably, we observe a new feature in the resistivity, that seems to require an explanation in terms of an electronic phase of unknown origin in U$_2$Rh$_3$Si$_5$, and which appears as a precursor phase of the magnetic transitions. After all, there seems to be a very close interdependence of electronic, magnetic and structural degrees of freedom in U$_2$Rh$_3$Si$_5$. Describing and understanding this electronic phase/precursor will be the main task of future studies of this and related materials. However, prior to a more detailed discussion of the unusual electronic behavior seen in the present set of experiments, we will first summarize and evaluate the apparently more common findings on the magnetic (and structural) properties
	
	Starting with the observations on the phase diagram of U$_2$Rh$_3$Si$_5$, for the first time various steps in the high field magnetization of the $a$ axis were observed, establishing the existence of a rather complex magnetic phase diagram for this crystallographic direction. In detail, we conclude that the ground state AFM phase I$_a$ transforms with a first-order transition into phase II$_a$. At higher fields, again there is first-order character at the transition from phase II$_a$ into the paramagnetic phase. As a subtlety, this transition even has double-step character, implying that there is a very narrow intermediate magnetic phase III$_a$. Correspondingly, phases II$_a$ and III$_a$ must be intermediate between the AFM alignment of phase I$_a$ and the polarized spin state of the paramagnetic high-field phase. Likely, there is additional staggering of magnetic moments in these field-induced phases as compared to the AFM phase I$_a$, for instance with a stacking of moments such as up-up-down etc., as discussed for instance for the staircase magnetization scenario in CeSb \cite{Rossat-Mignod1977}.
	
	For the $b$ axis at low temperatures the high field magnetization and magnetostriction show a single magnetic transition connected with a large volume change. This appears qualitatively to be consistent with the bootstrapping scenario \cite{Becker1997,Takeuchi1997,Feyerherm1997,Leisure2005}. But on top of this, the phase diagram for the $b$ axis exhibits various particularities. At temperatures $\leq 10$\,K and high fields the transition from the phase I$_b$ in the paramagnetic phase has a clear first order character demonstrated by sharp jumps in the magnetization and magnetostriction. However, for higher temperatures in these measurements the transition becomes broader and transforms into a two-step transition, this way defining an intermediate phase range I'$_b$. From their appearances, in this temperature and field range the magnetic transitions I$_b$ $\rightarrow$ I'$_b$ and I'$_b$ $\rightarrow$ PM have more of a second order or mixed phase character, while in fields $B \rightarrow 0$\,T and temperatures close to $T_N$ the step-like temperature dependence of the susceptibility again signals a first order transition. Finally, for the $c$ axis no field induced transitions have been recorded up to 65\,T, attesting to the very large magnetic anisotropy of this material.
	
	
	Over the years, a number of U-intermetallics, namely UNiAl \cite{Brueck1994}, UPt$_2$Si$_2$ \cite{schulze2012,Grachtrup2017}, UN \cite{Shrestha2017}, USb$_2$ \cite{Stillwell2017}, UIrSi$_3$ \cite{Valenta2018} and UIrGe \cite{Pospisil2018}, emerged with a similar change from a second order-like to a first order magnetic transition at low temperatures and high fields. It leads to tricritical points in the magnetic phase diagrams for these materials \cite{Grachtrup2017,Shrestha2017,Stillwell2017,Valenta2018,Pospisil2018}. In the following we will briefly compare the reports for these materials with $\mathrm{U_2Rh_3Si_5}$.
	
	The mentioned materials all have an AFM ground state with N\'{e}el temperatures between $16.5$\,K (UIrGe) and $202$\,K (USb$_2$) \cite{Brueck1994,Grachtrup2017,schulze2012,Shrestha2017,Stillwell2017,Valenta2018,Pospisil2017,Pospisil2018}. Starting with the field dependent magnetization, they have in common a sharp jump at the corresponding critical fields for low temperatures, indicative of a first order phase transition \cite{Brueck1994,schulze2012,Shrestha2017,Stillwell2017,Valenta2018,Pospisil2018}. With higher temperatures the transitions shift to lower fields and transform into an S-like shape indicating a more gradual transition into the polarized states \cite{schulze2012,Shrestha2017,Stillwell2017,Valenta2018,Pospisil2018}. This is similar to our measurements on $\mathrm{U_2Rh_3Si_5}$ as shown in Fig. \ref{magnetization}. But there are also differences visible, \textit{i.e.}, in the low field susceptibility. While $\mathrm{U_2Rh_3Si_5}$ shows a sharp jump at the critical temperature (see Fig. \ref{susceptibility}), the other materials exhibit a smooth rise of the susceptibility up to the critical temperature typical for the second order phase transition in common antiferromagnets \cite{Brueck1994,schulze2012,Valenta2018,Pospisil2018,Samsel-Czekala2007,Wawryk2006}. Only for UIrGe, in Ref. \cite{Pospisil2018} the temperature dependent susceptibility, measured in different magnetic fields, sharpens up for higher fields. This was interpreted as a change of the transition from second to first order. The susceptibility of $\mathrm{U_2Rh_3Si_5}$ resembles the high field behavior of UIrGe, reflecting that at $T_N$ $\mathrm{U_2Rh_3Si_5}$ exhibits a first order transition. In summary, regarding the magnetic properties, there are various U-systems with tricritical points in the magnetic phase diagram showing similarities to $\mathrm{U_2Rh_3Si_5}$. Especially, the change of the transition in the magnetization to low temperatures indicates that $\mathrm{U_2Rh_3Si_5}$ may exhibit tricritical points in the phase diagram for the $a$ and $b$ axes. 
	
	Regarding the interpretation of the first order antiferromagnetic transition in $\mathrm{U_2Rh_3Si_5}$ being the result of a bootstrapping effect \cite{Becker1997,Wang1968,Wang1969}, the question arises if based on our experiments we can draw conclusions as to the validity of the argument. Globally, our experimental findings appear to be consistent with a bootstrapping scenario: The magnetic properties of $\mathrm{U_2Rh_3Si_5}$ would classify this material as a uranium intermetallic with well-localized $f$ electrons. In principle, a description of the $f$ states within a conventional crystal electric field scheme appears possible and has been proposed before \cite{Leisure2005}. Hence, it should also be possible to extend the modelling to include the field-dependent effects that we report here. Unfortunately, the low crystallographic symmetry and complex magnetic structure, together with the very pronounced local anisotropy of the $f$ electrons, complicate matters to the effect that a detailed and quantitatively accurate modelling of the various thermodynamic properties still appears not attainable. Thus, while being consistent with the bootstrapping scenario, at present our experimental data do not yield definite proof of it. 
		
	Aside from the magnetic transitions, the upturn in the resistivity of $\mathrm{U_2Rh_3Si_5}$ for $B || a$ and $b$ axis and corresponding drop along the $c$ axis at a temperature $T^* > T_N$ denotes yet another phase transition. It is observed in the resistivity but does not show a signature in the susceptibility or the structural parameters. Therefore, it is not a magnetic transition, {\it i.e.}, an ordering transition in spin space. Instead, the anomalies in the resistivity signal an electronic phase transition, resulting in a change of the carrier density or the scattering cross section due to band structure modifications. We note that for such electronic transitions it is a common occurrence to be clearly observable only in experimental probes such as electronic transport, but not in thermodynamic quantities.
	
	Compared to $\mathrm{U_2Rh_3Si_5}$, the resistivities of UNiAl \cite{Brueck1994}, UN \cite{Samsel-Czekala2007} and UIrGe \cite{Pospisil2017} show some interesting similarities close to $T_N$. The resistivity of the $c$ axis of UNiAl exhibits a peak like anomaly comparable to the feature in $\mathrm{U_2Rh_3Si_5}$. Unfortunately, from the data shown in Ref. \cite{Brueck1994} it can not be assessed if the magnetic transition corresponds to the upturn or the maximum of the resistivity. For a better comparison with $\mathrm{U_2Rh_3Si_5}$ a more accurate and field dependent measurement of the resistivity is necessary. In UN and UrIrGe similar structures in the resistivity are visible \cite{Samsel-Czekala2007,Pospisil2017}. For these materials, it is established experimentally that the rise of the resistivity is connected to the AFM transition temperature, and not the maximum. Here, too, it would be interesting to investigate in more detail the interplay of the electronic and magnetic behavior and compare it to $\mathrm{U_2Rh_3Si_5}$. Altogether, based on our results on $\mathrm{U_2Rh_3Si_5}$, it seems worthwhile to reinvestigate UNiAl, UN and UIrGe, to verify that the features in the resistivity truly correspond to an AFM transition.
	
	Conversely, we note that the electronic phase seen in zero magnetic field at $T^* > T_N$ sets $\mathrm{U_2Rh_3Si_5}$ apart from all other materials. Qualitatively, the zero field resistive behavior at $T^*$ does have a resemblance to a charge density wave as seen for example in $\mathrm{Lu_5Ir_4Si_{10}}$ \cite{Becker1999a,Jung2003}. Only, in $\mathrm{U_2Rh_3Si_5}$, there is a very strong field dependence of the transition temperature $T^*$: It evolves in a fashion similar to antiferromagnetic transitions, as it shifts to lower temperatures with magnetic field. For a charge density wave one would not expect such a field dependence, implying that an explanation of the observations along these lines appears impossible. Hence, the anomaly at $T^*$ denotes an electronic state in $\mathrm{U_2Rh_3Si_5}$ of as yet unknown type.
	
	Remarkably, the field dependence of the phase border line of $T^*$ is even somewhat stronger than that of the AFM phases. This results in a coexistence regime of electronic and magnetically ordered phases. Specifically, we have observed that the novel electronic phase transition in $\mathrm{U_2Rh_3Si_5}$ merges with the magnetic phase transitions. This finding raises questions about our above statement of well-localized $f$ electrons in this compound. At around the merger of the phase border lines of $T^*$ and I$_b/$I'$_b$ the overall behavior of the corresponding physical properties appears to be more in line with a second order phase transition. Notably, in this range the magnetization transition smears out and does not have such a pronounced local moment character as it has at low temperatures. It appears as if the coexistence of local moment magnetism with an (itinerant?) electronic phase weakens the local magnetic moment character. Conversely, the observed strong field dependence of the transition temperature of the electronic phase would then still attest to the residual local moment character of the U ions.
	
	This observation of course directly relates to the issue of the proper description of uranium $f$ magnetic moments, which can have localized, itinerant or even dual character. A bootstrapping scenario in the presence of a dual character of the $f$ electrons would necessarily lead to a complex interplay of local moment physics and band structure effects. We speculate that the subtle balance of electronic and magnetic phases observed in $\mathrm{U_2Rh_3Si_5}$ may reflect such a scenario. In this situation, experimentally what is required as a next step (and is notoriously hard to attain in uranium compounds) would be experimental information about the local moment character of the uranium $f$ electrons and the band structure. Additionally, the possible electric transition should be investigated by further electrical measurements like Seebeck effect, Hall effect or thermal conductivity. Moreover, experiments to higher fields for the $a$ axis such as resistivity might provide additional insight into this complex topic.

	\begin{acknowledgments}
		The crystal has been grown and characterized in the Materials Growth and Measurement Laboratory MGML which is supported within  the  program  of  Czech  Research Infrastructures (Project No. LM2018096).  A portion of this work was performed at the National High Magnetic Field Laboratory, which is supported by the National Science Foundation Cooperative Agreement No. DMR-1644779 and the state of Florida. S. S. acknowledges support by the Magnet Lab. Visiting Scientist Program of the NHMFL.
	\end{acknowledgments}


\begin{thebibliography}{00}
		\bibitem{Pfleiderer2009} C. Pfleiderer, Rev. Mod. Phys. {\bf 81}, 1551 (2009).
		\bibitem{Brando2016} M. Brando, D. Belitz, F. M. Grosche and T. R. Kirkpatrick, Rev. Mod. Phys. {\bf 88}, 025006 (2016)
		\bibitem{Mydosh2017} J.A. Mydosh, Adv. Phys. {\bf 66}, 263 (2017)
		\bibitem{Faber1976} J. Faber and G. H. Lander, Phys. Rev. B {\bf 14}, 1151 (1976)
		\bibitem{Giannozzi1987} P. Giannozzi and P. Erd\"os, J. Magn. Magn Mat. {\bf 67}, 75 (1987)
		\bibitem{Santini2009} P. Santini, S. Carretta, G. Amoretti, R. Caciuffo, N. Magnani and G. H. Lander, Rev. Mod. Phys. {\bf 81}, 807 (2009)
		\bibitem{Jaime2017} M. Jaime, A. Saul, M. Salamon, V. S. Zapf, N. Harrison, T. Durakiewicz, J. C. Lashley, D. A. Andersson, C. R. Stanek, J. L. Smith and K. Gofryk, Nat. Commun. {\bf 8}, 99 (2017)
		\bibitem{Andres1978} K. Andres, D. Davidov, P. Dernier, F. Hsu, W. A. Reed and G. J. Nieuwenhuys, Solid State Commun. {\bf 28}, 405 (1978)
		\bibitem{McEwen1993} K. A. McEwen, U. Steigenberger and J. L. Martinez, Physica B {\bf 186-188}, 670 (1993)
		\bibitem{Walker2006} H. C. Walker, K. A. McEwen, D. F. McMorrow, S. B. Wilkins, F. Wastin, E. Colineau and D. Fort, Phys. Rev. Lett. {\bf 97}, 137203 (2006)
		\bibitem{Becker1997} B. Becker, S. Ramakrishnan, A. A. Menovsky, G. J. Nieuwenhuys and J. A. Mydosh, Phys. Rev. Lett. {\bf 78}, 1347 (1997)
		\bibitem{Takeuchi1997} T. Takeuchi, T. Yamada, Y. Miyako, K. Oda, K. Kindo, B. Becker, S. Ramakrishnan, A. A. Menovsky, G. J. Nieuwenhuys and J. A. Mydosh, Phys. Rev. B {\bf 56}, 10778 (1997)
		\bibitem{Feyerherm1997} R. Feyerherm, C. R. Wiebe, B. D. Gaulin, M. F. Collins, B. Becker, R. W. A. Hendrikx, T. J. Gortenmulder, G. J. Nieuwenhuys and J. A. Mydosh, Phys. Rev. B {\bf 56}, 13693 (1997)
		\bibitem{Leisure2005} R. G. Leisure, S. Kern, F. R. Drymiotis, H. Ledbetter, A. Migliori and J. A. Mydosh, Phys. Rev. Lett. {\bf 95}, 075506 (2005)
		\bibitem{Wang1968} Y.-L. Wang and B. R. Cooper, Phys. Rev. {\bf 172}, 539 (1968)
		\bibitem{Wang1969} Y.-L. Wang and B. R. Cooper, Phys. Rev. {\bf 185}, 696 (1969)
		\bibitem{Hickey1990} E. Hickey, B. Chevalier, P. Gravereau and J. Etourneau, J. Magn. Magn. Mater {\bf 90-91}, 501 (1990)
		\bibitem{Galli2000} F. Galli, R. Feyerherm, K. Prokes and G.J. Nieuwenhuys, Physica B {\bf 276-278}, 632 (2000)
		\bibitem{Haga1998} Y. Haga, T. Honma, E. Yamamoto, H. Ohkuni, Y. {\={O}}nuki, M. Ito and N. Kimura, Jpn. J. Appl. Phys. {\bf 37}, 3604 (1998)
		\bibitem{Daou2010} R. Daou, F. Weickert, M. Nicklas, F. Steglich, A. Haase and Mathias Doerr, Rev. Sci. Instrum. {\bf 81}, 033909 (2010)
		\bibitem{Jaime2012} M. Jaime, R. Daou, S. A. Crooker, F. Weickert, A. Uchida, A. E. Feiguin , C. D. Batista, H. A. Dabkowska and B. D. Gaulin, PNAS {\bf 109}, 12404 (2012)
		\bibitem{Jaime2017a} M. Jaime, C. C. Moya, F. Weickert, V. Zapf, F. Balakirev, M. Wartenbe, P. Rosa, J. Betts, G. Rodriguez, S. Crooker and Ramzy Daou, Sensors {\bf 17}, 2572 (2017)
		\bibitem{Detwiler2000} J. A. Detwiler, G. M. Schmiedeshoff, N. Harrison, A. H. Lacerda, J. C. Cooley and J. L. Smith, Phys. Rev. B {\bf 61}, 402 (2000)
		\bibitem{Becker1999} B. Becker, A. C. Wallast, J. A. A. J. Perenboom, G. J. Nieuwenhuys and J.A Mydosh, Physica B {\bf 259-261}, 250 (1999)
		\bibitem{Mackintosh1962} A. R. Mackintosh, Phys. Rev. Lett. {\bf 9}, 90 (1962)
		\bibitem{Elliott1963} R. J. Elliott and F. A. Wedgwood, Proc. Phys. Soc. {\bf 81}, 846 (1963)
		\bibitem{Hegland1963} D. E. Hegland, S. Legvold and F. H. Spedding, Phys. Rev. {\bf 131}, 158 (1963)
		\bibitem{Mentink1996} S. A. M. Mentink, T. E. Mason, S. S\"ullow, G. J. Nieuwenhuys, A. A. Menovsky, J. A. Mydosh and J. A. A. J. Perenboom, Phys. Rev. B {\bf 53}, R6014 (1996)
		
	\bibitem{newDis} Close and comparative inspection of the AFM transition as seen in susceptibility and resisitivity indicates that the value $T_N(\rho)$ is consistently slightly higher than $T_N(\chi)$. We believe that this reflects the first order nature of the phase transition, leading to coexistence of paramagnetic and antiferromagnetic phase volume. With the drop of the resistivity in the AFM phase, as a transport technique the resistivity will be more sensitive to the onset of magnetic order, as it will probe the percolative resistance path provided by the low resistive AFM phase. In contrast, susceptibility and magnetostriction will more closely reflect the phase volume of the AFM phase, leading to an effectively smaller $T_N$ seen with these techniques.
		
		\bibitem{Rossat-Mignod1977} J. Rossat-Mignod, P. Burlet, J. Villain, H. Bartholin, W. Tcheng-Si, D. Florence and O. Vogt, Phys. Rev. B {\bf 16}, 440 (1977)
		\bibitem{Brueck1994} E. Br\"uck, H. Nakotte, F. R. de Boer, P. F. de Ch{\^{a}}tel, H. P. van der Meulen, J. J. M. Franse, A. A. Menovsky, N. H. Kim-Ngan, L. Havela, V. Sechovsky, J. A. A. J. Perenboom, N. C. Tuan and J. Sebek, Phys. Rev. B {\bf 49}, 8852 (1994)
		\bibitem{schulze2012} D. Schulze Grachtrup, M. Bleckmann, B. Willenberg, S. S\"ullow, M. Bartkowiak, Y. Skourski, H. Rakoto, I. Sheikin and J. A. Mydosh, Phys. Rev. B {\bf 85}, 054410 (2012)
		\bibitem{Grachtrup2017} D. Schulze Grachtrup, N. Steinki, S. S\"ullow, Z. Cakir, G. Zwicknagl, Y. Krupko, I. Sheikin, M. Jaime and J. A. Mydosh, Phys. Rev. B {\bf 95}, 134422 (2017)
		\bibitem{Shrestha2017} K. Shrestha, D. Antonio, M. Jaime, N. Harrison, D. S. Mast, D. Safarik, T. Durakiewicz, J.-C. Griveau and K. Gofryk, Sci. Rep. {\bf 7}, 6642 (2017)
		\bibitem{Stillwell2017} R. L. Stillwell, I-L. Liu, N. Harrison, M. Jaime, J. R. Jeffries and N. P. Butch, Phys. Rev. B {\bf 95}, 014414 (2017)
		\bibitem{Valenta2018} J. Valenta, F. Honda, M. Vali{\v{s}}ka, P. Opletal, J. Ka{\v{s}}til, M. M{\'{\i}}{\v{s}}ek, M. Divi{\v{s}}, L. Sandratskii, J. Prchal and V. Sechovsk{\'{y}}, Phys. Rev. B {\bf 97}, 144423 (2018)
		\bibitem{Pospisil2018} J. Posp{\'{\i}}{\v{s}}il, Y. Haga, Y. Kohama, A. Miyake, S. Kambe, N. Tateiwa, M. Vali{\v{s}}ka, P. Proschek, J. Prokle{\v{s}}ka, V. Sechovsk{\'{y}}, M. Tokunaga, K. Kindo, A. Matsuo and E. Yamamoto, Phys. Rev. B {\bf 98}, 014430 (2018)
		\bibitem{Pospisil2017} J. Posp{\'{\i}}{\v{s}}il, J. Gouchi, Y. Haga, F. Honda, Y. Uwatoko, N. Tateiwa, S. Kambe, S. Nagasaki, Y. Homma and E. Yamamoto, J. Phys. Soc. Jpn {\bf 86}, 044709 (2017)
		\bibitem{Samsel-Czekala2007} M. Samsel-Czeka{\l}a, E. Talik, P. de V. Du Plessis, R. Tro{\'{c}}, H. Misiorek and C. Su{\l}kowsk, Phys. Rev. B {\bf 76}, 144426 (2007)
		\bibitem{Wawryk2006} R. Wawryk, Philos. Mag. {\bf 86}, 1775 (2006)
		\bibitem{Becker1999a} B. Becker, N. G. Patil, S. Ramakrishnan, A. A. Menovsky, G. J. Nieuwenhuys, J. A. Mydosh, M. Kohgi and K. Iwasa, Phys. Rev. B {\bf 59}, 7266 (1999)
		\bibitem{Jung2003} M. H. Jung, H. C. Kim, A. Migliori, F. Galli and J. A. Mydosh, Phys. Rev. B {\bf 68}, 132102 (2003)		
	\end{thebibliography}
\end{document}